\documentclass[sigconf]{acmart}

\usepackage{float}
\usepackage{booktabs}
\usepackage{tabularx}
\usepackage[english]{babel}
\usepackage[capitalize, noabbrev]{cleveref}

\usepackage{graphicx, caption, subcaption}
\usepackage{todonotes}

\usepackage[export]{adjustbox}

\AtBeginDocument{%
  \providecommand\BibTeX{{%
    \normalfont B\kern-0.5em{\scshape i\kern-0.25em b}\kern-0.8em\TeX}}}

\newcommand{\hlidea}[1]{\textit{#1}}

\definecolor{DanielsColor}{rgb}{0.9,0.6,0.1}

\definecolor{HaisColor}{rgb}{0.9,0.3,0.9}

\definecolor{FloriansColor}{rgb}{0,0.3,0.9}

\definecolor{LukasColor}{rgb}{0.3,0.9,0.3}

\setcopyright{rightsretained}
\copyrightyear{2022}
\acmYear{2022}

\acmConference[CHI'22 Workshops]{CHI'22 Workshops}{May 10, 2022}{New Orleans, LA}
\acmBooktitle{GenAICHI: Generative AI and Computer Human Interaction, Workshop at CHI'22}

\begin{document}

\title{How to Prompt? Opportunities and Challenges of Zero- and Few-Shot Learning for Human-AI Interaction in Creative Applications of Generative Models}

\renewcommand{\shorttitle}{How to Prompt? Opportunities and Challenges of Zero- and Few-Shot Learning for Interaction}

\author{Hai Dang}
\email{hai.dang@uni-bayreuth.de}
\affiliation{%
  \institution{Research Group HCI + AI, Department of Computer Science, University of Bayreuth}
  \city{Bayreuth}
  \country{Germany}
}

\author{Lukas Mecke}
\email{lukas.mecke@unibw.de}
\affiliation{%
  \institution{Bundeswehr University Munich}
  \city{Munich}
  \country{Germany}
}
\affiliation{%
  \institution{LMU Munich}
  \city{Munich}
  \country{Germany}
}
\author{Florian Lehmann}
\email{florian.lehmann@uni-bayreuth.de}
\affiliation{%
  \institution{Research Group HCI + AI, Department of Computer Science, University of Bayreuth}
  \city{Bayreuth}
  \country{Germany}
}
\author{Sven Goller}
\email{sven.goller@uni-bayreuth.de}
\affiliation{%
  \institution{Research Group HCI + AI, Department of Computer Science, University of Bayreuth}
  \city{Bayreuth}
  \country{Germany}
}

\author{Daniel Buschek}
\orcid{0000-0002-0013-715X}
\email{daniel.buschek@uni-bayreuth.de}
\affiliation{%
  \institution{Research Group HCI + AI, Department of Computer Science, University of Bayreuth}
  \city{Bayreuth}
  \country{Germany}
}

\renewcommand{\shortauthors}{Dang, et al.}

\begin{abstract}
Deep generative models have the potential to fundamentally change the way we create high-fidelity digital content but are often hard to control. \textit{Prompting} a generative model is a promising recent development that in principle enables end-users to creatively leverage zero-shot and few-shot learning to assign new tasks to an AI ad-hoc, simply by writing them down. However, for the majority of end-users writing effective prompts is currently largely a trial and error process. To address this, we discuss the key opportunities and challenges for interactive creative applications that use prompting as a new paradigm for Human-AI interaction. Based on our analysis, we propose four design goals for user interfaces that support prompting. We illustrate these with concrete UI design sketches, focusing on the use case of creative writing. The research community in HCI and AI can take these as starting points to develop adequate user interfaces for models capable of zero- and few-shot learning.
\end{abstract}

\ccsdesc[500]{Computing methodologies~Cooperation and coordination}
\ccsdesc[500]{Human-centered computing~Collaborative interaction}

\keywords{HCI, Artificial Intelligence, Co-Creation, UI Design, Prompt Engineering}

\maketitle

\section{Introduction}

Recent advances in deep generative systems have enabled the creation of high fidelity media such as music, images, and text. These developments are becoming increasingly relevant also for creative domains as they potentially enable more collaborative and dynamic co-creative work of humans and AI systems. %
Trained on a large data corpus, such generative system can now synthesize new original content. However, user control over the generated output remains challenging. This requires effective interactions and user interfaces, as well as underlying AI mechanisms to support conditional and controlled generation.

One emergent interaction technique, especially for text generation, is the use of natural language \textit{prompts} to steer the generative system: %
It was found that large language models (LLMs) do not necessarily need to be fine-tuned for specific tasks, such as for sentiment classification or topical text generation~\cite{Brown2020gpt3}. Instead, users may write a short text prompt to tell the system what to generate. Depending on how many examples are provided in the text prompt, we may differentiate between zero-, one-, and few-shot learning. Zero shot-learning refers to prompts where no specific example is given, such as: ``Translate the following sentence into German: Good morning, how are you doing?''. However, providing one example (one-shot learning) or more examples (few-shot learning) in the prompt may improve the output. For a detailed overview, see the recent survey by \citet{liu_pre_train_2021}.

From an end-user perspective, writing prompts is largely approached as a trial and error process. Several influences have been identified and addressed in research (e.g. order of words) and best practices are shared informally on the web. However, the design of an effective prompting interface has not yet been explored systematically from a Human-Computer Interaction perspective. To address this, we collected and present key opportunities and challenges for prompting as a new paradigm for Human-AI interaction %
and propose and illustrate four design goals for user interfaces that support this. With this, we provide an overview and starting points for future research on user interfaces for models capable of zero- and few-shot learning.

\section{Related work}
We give an overview of work on interactive use of prompting, as well as techniques developed in prompt engineering.

\subsection{Interactive Applications Using Few-shot Learning}\label{sec:rw_interactive}
Writing effective prompts is often not straightforward for users. A couple of recent projects have addressed this:
For example, \textit{AI Chains} is an interactive tool to combine a sequence of (sub-)tasks to be solved with repeated runs of one LLM \cite{wu2021ai}. Related, \textit{SynthBio} uses few-shot learning to synthesise a data corpus of biographies~\cite{yuan2021synthbio} . This text synthesis consists of multiple steps, involving both an LLM to synthesize new biographies and humans to rate them. Both these systems \cite{wu2021ai, yuan2021synthbio} focus on a combination of prompts as a means to reduce task complexity. In this paper, we highlight the idea of supporting prompt composition and combination as a key design goal and direction for user interfaces for prompting.

Further interactive examples include \textit{Story Centaur} by \citet{pietrzak2021}, an application for non-technical users to quickly construct prompts with few-shot examples: Through a graphical user interface users can define input and output text phrases, as well as phrases to be included before, in-between, or after each such example. The tool formats these components as a text prompt for few-shot learning. %
Related, \citet{austin2021program} also built a tool that provides a template for prompt input in its UI, here to support synthesis of computer programs. %
Supporting prompt formulation and input is one of the key design goals for user interfaces that we discuss in this paper.

\subsection{Prompt Engineering}

Prompt engineering refers to the systematic practice of constructing prompts to improve the generated output of a generative model. Particular examples include these insights: \citet{lu2021fantastically} have found that word order affects the generated outcome. 
Moreover, while providing longer context in prompts empirically resulted in better text outputs \cite{khandelwal2018sharp, beltagy2020longformer}, \citet{wu2021ai} have found that task instructions within a longer context body can also conflict with each other. \citet{o2021context} provide a framework to determine ``how much'' information of the original context the prompt can actively use in the generation process. 
Another approach to elicit better prompts is to use the LLM itself for problem elaborations \cite{betz2021thinking}, inspired by the ``think aloud'' strategy humans use to solve complex problems. 

Other strategies improve text prompts algorithmically: For example, \citet{liu2021makes} use sentence embedding on prompts to automatically sample neighboring prompt alternatives for comparison. \textit{AutoPrompt}~\cite{shin2020eliciting} optimises a set of words in the prompt using the gradient with respect to a task output (e.g. sentiment classification). Related, ``prefix tuning'' draws inspiration from prompting but optimises context tokens that do not need to map to actual words~\cite{li2021prefixtuning}.

Finally, prompt-style input/output can also be used to finetune language models (``Pattern Exploiting Training''~\cite{schick_exploiting_2021}) yet this is different from the prompts we discuss in this paper in that we aim to utilise their capability to perform (end user-specified) tasks ad-hoc without task-specific model training/finetuning.

\section{Method}

We organised two one-hour long brainstorming sessions with five HCI researchers to discuss and collect speculative opportunities and challenges that result from the prompt design paradigm to spark new discussions about the impact of prompt writing for creative expression. The first session included multiple phases. First, each participant quietly noted potential opportunities and challenges on a virtual whiteboard. In the second step, each participant briefly introduced their idea and then similar ideas were clustered. Finally, following a discussion and using the previously formed clusters, we identified overarching topics that we introduce in sections \ref{sec:opportunities} and \ref{sec:challenges}.

In the second session we started with a similar approach and collected ideas on the interactive use of prompt writing. In the brainstorming session we listed existing interactive systems that support users to prompt large language models (refer to \ref{sec:rw_interactive}) and drew inspiration for designing new scenarios to help users in their creative writing process. We started with the question of what creative writing means, followed by each researcher noting down potential goals and motivations for creative writing. Based on these notes we identified a three phase creative writing process which the ideal prompt writing support tool should cover at the very least (\ref{sec:creative_writing}).

\section{Opportunities}\label{sec:opportunities}

We highlight three opportunities that prompting may bring to the table. While our concrete examples focus on the use case of creative writing, the opportunities themselves are not limited to that domain.

\subsection{End-User Programming of Creative Tools}
Fundamentally, prompting LLMs at runtime gives the user the opportunity to \hlidea{define new tools ad hoc as needed}, without retraining the model. This could be especially interesting for users without a technical background, because they can \hlidea{use natural language} to define the tools they need and when they need it. 

By defining generic prompts the users may also build up a library of generative tools for \hlidea{reuse in different contexts}. Defining such tools can be done through \hlidea{programming by example} (few-shot learning) or \hlidea{by declaration} (zero shot learning).

\hlidea{Generating code} from natural language prompts (e.g. OpenaAI's \textit{Codex}~\cite{chen2021evaluating}) empowers users to create, adapt and appropriate digital tools (cf.~\cite{Klokmose2015}). However, prompting need not be limited to an interpretation as ``code generation''; it could also be viewed as powering a \hlidea{conversation} of humans and AI involved in a task, in which users can now define and \hlidea{delegate tasks verbally}. This elevates the AI to be an active and dynamic contributor for the joint generation of new media instead of being a predefined assistant or toolbox. 

\subsection{Extending and Augmenting Creative Expressiveness}

It is often not straight forward to sufficiently describe certain concepts in detail, especially in artistic contexts. For example, it would be very difficult to exactly describe certain text styles, but it is \hlidea{easy to express vague concepts} via a prompt to an LLM, such as to ``write a text in the prose of Shakespeare''. Other creative use involves \hlidea{prompting across modalities}, such as text to image models~\cite{ramesh2021zeroshot}, or prompting with images from which the generative systems may extract the mood or atmosphere (cf. AI-supported moodboards~\cite{Koch2020}). %
Formatting and rephrasing prompts have been shown to influence the text outcome. Therefore, prompting may allow users to \hlidea{synthesize new forms and styles}, or translate between different ones. For example, a prompt might be designed to turn a novel into a screen play text or a non-linear format (e.g. text adventure).

\subsection{Providing Inspiration and Feedback}

Prompts may help to \hlidea{overcome a creative block} by generating content. Related, prompts may be useful to simulate \hlidea{personal feedback} on a given text (e.g. from the perspective of a famous author). %
Users can also prompt the model to provide \hlidea{input based on other works} and styles (e.g. ``What would Shakespeare write here?''), or more generally to \hlidea{optimise drafts} with various objectives (e.g. ``Make this text more formal''). 
Moreover, creating new media, such as writing an article, typically involves iteration. Prompts may help users to \hlidea{quickly iterate} and \hlidea{explore multiple variations} of a text, also by using different examples and by \hlidea{giving high-level directions} (e.g. keywords, styles). %

\section{Challenges}\label{sec:challenges}

Here we describe key challenges of interactive use of prompting. These are based on the literature, material shared by practitioners online, and our own experiences. %

\subsection{Trial and Error / Lack of Guidance}

Prompt design or engineering has seen little systematic research from an HCI perspective and for users remains mostly a \hlidea{trial and error} process. This is also evident from the best practices shared in informal resources such as blogs and forums on the internet~\cite{branwen2020blogpost, cohere2021blog}. Beyond that, recent work in HCI and AI reports on techniques such as elaboration~\cite{betz2021thinking}, optimisation of key ``trigger'' words~\cite{shin2020eliciting} or changing the word order~\cite{lu2021fantastically}. 
A collection of prompts has also been started, with a tool to support researchers in writing prompts on datasets~\cite{sanh2021multitask}\footnote{\url{https://github.com/bigscience-workshop/promptsource}, \textit{last accessed 27.01.2022}}.
However, few of the practically discovered or researched strategies have led to dedicated user interfaces and interactive tools for non-technical users. 

The most elaborate user interface design we found is \textit{AI Chains}, which embodies the strategy of breaking down complex prompts into subtasks that are easier for the model~\cite{wu2021ai}. 
Extrapolating these observations raises questions about how future interfaces may support even more \hlidea{complex prompts} and \hlidea{exploring multiple prompts}. 
Overall, we envision that future interactive tools support users in creating and applying more effective prompts more efficiently and creatively. 

\subsection{Representation of Tasks and Effects / Prompt IO}

If prompts enable users to define custom AI tools, default graphical representations (e.g. icons) might not suffice to describe them. Thus, it is an open question how prompts are \hlidea{represented in the interface} and \hlidea{how users can interact} to apply or execute them. Realising this requires \hlidea{discoverability and explanations} as well: For instance, how might users be enabled to understand or remember what a prompt-based tool does without rereading the underlying prompt? 

Related, current prompts mainly \hlidea{incentivise verbal thinking} despite visuals or tacit knowledge and skills being part of many creative practices. It is not clear how to meaningfully incorporate images in prompts or how to visualise prompts (e.g. to integrate them into a GUI). 

In addition, prompts are often \hlidea{conceptualised as one-time executions} (e.g. write the prompt, then ``run'' it). However, creative work is often an iterative approach that requires the exploration of multiple potential outcomes and effects. Currently, it is an open question how interfaces might support such iterative explorations of input and output mappings for prompting. 

Moreover, current \hlidea{constraints may defeat the upside of using natural language}, for instance, if the prompt has to follow a strict structural format or specific ``hidden rules'' to be effective. Without adequate interactions and user interfaces, for creative practitioners, writing prompts might end up being very similar to writing code.

\subsection{Computational Costs, Generalisation and Ethical Concerns}
Large generative systems may introduce a \hlidea{noticeable delay} which might affect users' (creative) work process. For example, delays may prevent quick iterations. %
A related concern is \hlidea{gatekeeping} access to such tools if they are only feasible to operate, for example, for large companies. %

It is also not clear how well prompts \hlidea{generalise across different models}. Optimizing a prompt in the context of one model does not guarantee the same performance when using another generative system. Besides usability issues this may create a lock-in effect where users are forced to continue to use one specific system. %

Finally, content created via prompting is expected to replicate \hlidea{biases} known from large language models in general, such as racist and stereotypical language.

\section{Designing User Interfaces that Support Prompting}\label{sec:creative_writing}

\begin{figure*}[t]
    \centering
    \includegraphics[width=0.7\textwidth]{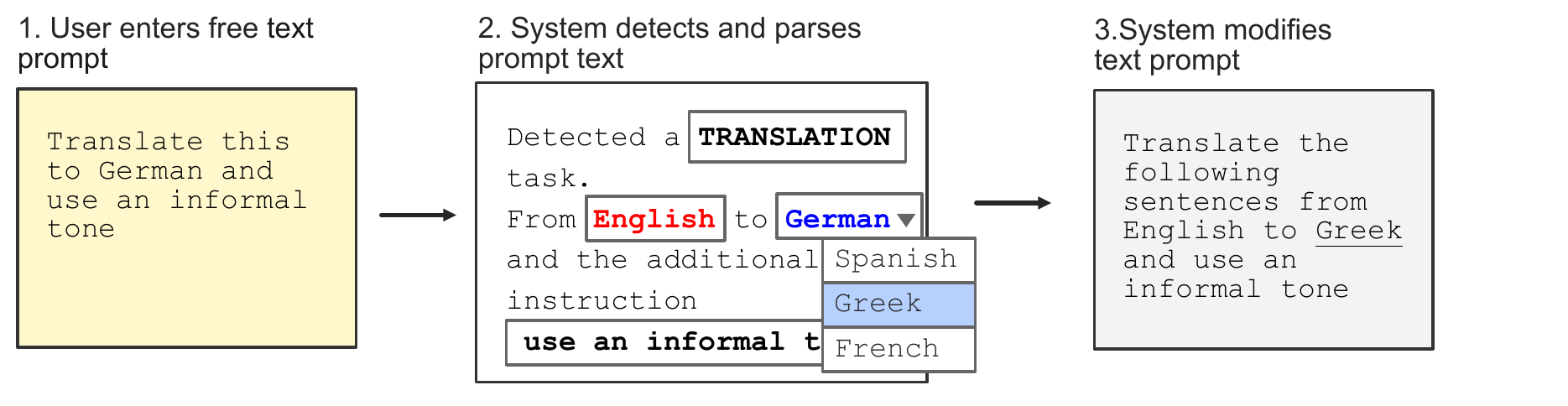}
    \caption{In this GUI example, users enter a prompt in natural language and the system automatically parses this input. Key parameters such as detected task and task parameters can thus be edited directly. Then, the system automatically creates the (refined) text prompt for the generative model.}
    \label{fig:task_parsing}
\end{figure*}

\begin{figure*}[t]
    \centering
    \includegraphics[width=0.7\textwidth]{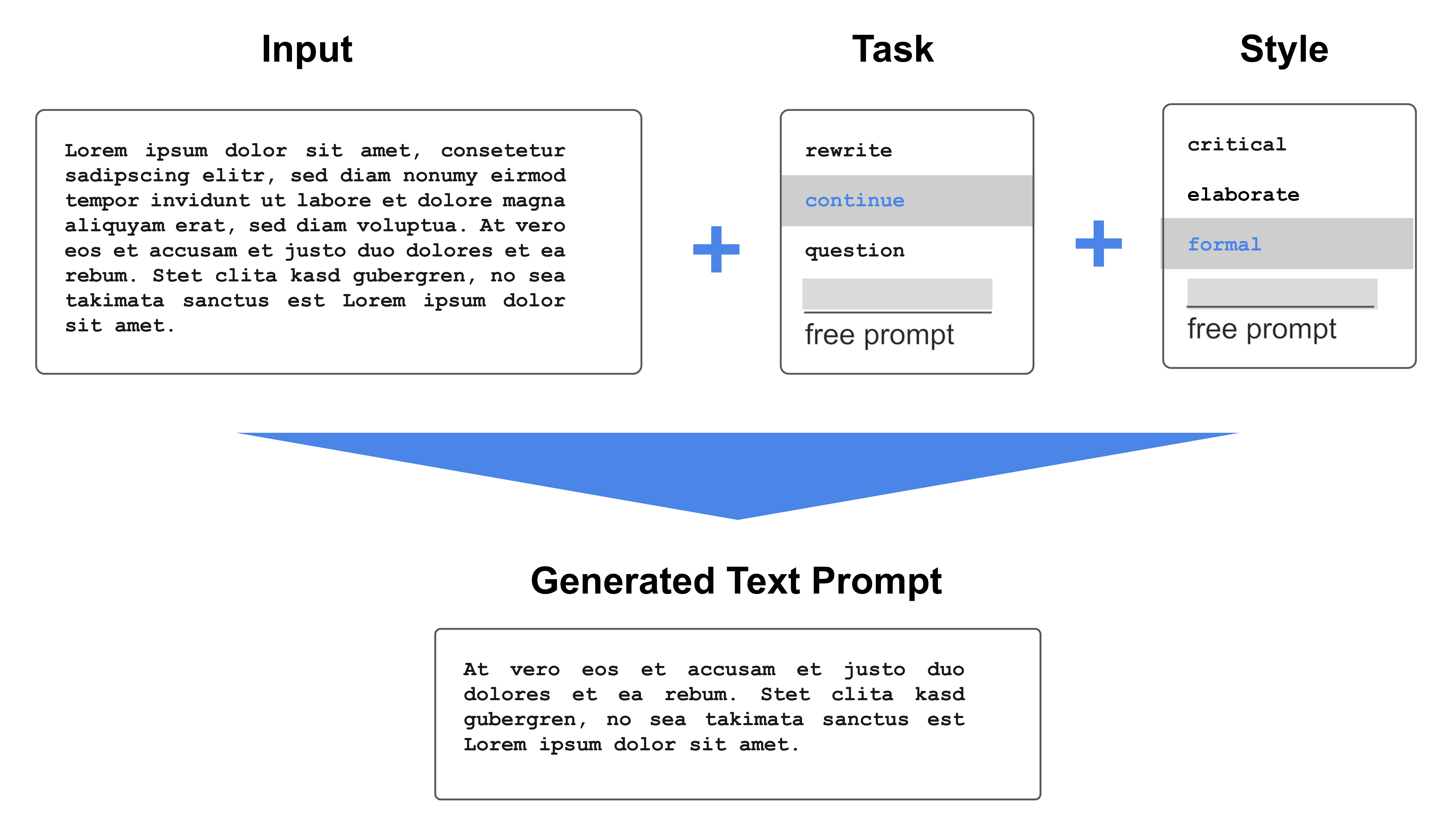}
    \caption{In this example GUI, users create a prompt not by writing from scratch but by selecting from predefined ``building blocks'' that have been proven to work well and cover typical use case-specific aspects. Free entry is still supported as well.}
    \label{fig:meta_prompts}
\end{figure*}

Based on the literature and own ideas collected in a design session, we have identified four design goals for user interfaces that support prompting. A particular design may also address multiple of these goals. We mainly discuss the use case of creative writing in our examples here but expect these design ideas to be useful starting points for future work more generally as well.

\subsection{Supporting Users in Formulating Prompts}
An effective user interface should support users in drafting prompts quickly and effortlessly. One way this could be achieved is to let the system automatically parse a user's input and create text prompts as in Fig. \ref{fig:task_parsing}. Furthermore, detected elements can be made editable as graphical UI elements (e.g. dropdown lists). While potentially easy to use, automatic detection may come with limitations, such as on the number of elements the systems recognises. A bottom-up approach would allow a user to manually define prompt improvements by selecting pre-defined parameters. This way, users have more granular control over the text prompt (see Fig. \ref{fig:meta_prompts}).

\subsection{Supporting Users in Combining Prompts}
\begin{figure*}[t]
    \centering
    \includegraphics[width=\textwidth]{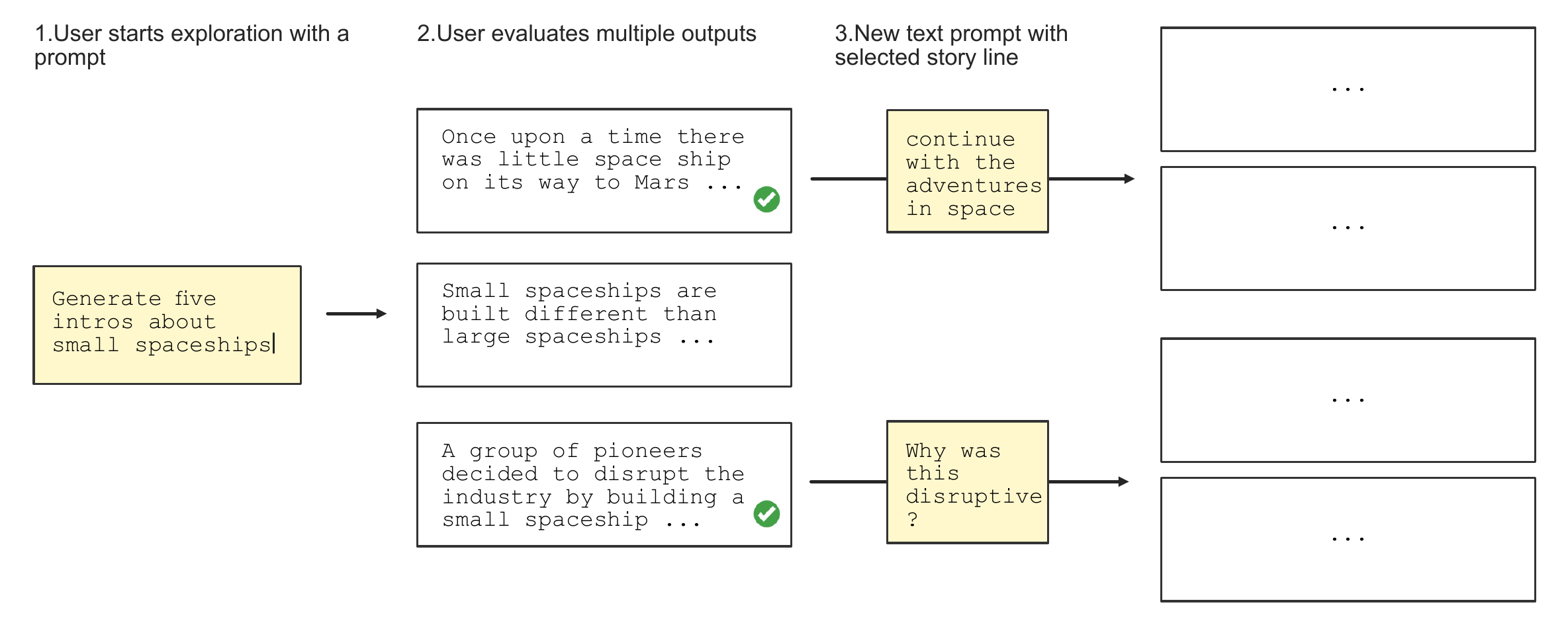}
    \caption{This example GUI focuses on prompt exploration and combination: Users write prompts to direct a ``narrative tree'' showing multiple possible responses to each prompt. Users select some of them as context for the next prompt(s), which direct the narrative further.}
    \label{fig:multiple_story_lines}
\end{figure*}

Another key aspect for interaction is to support users in combining multiple prompts \cite{wu2021ai}. For example, for creative writing, this could enable users to explore multiple story lines in parallel (see Fig. \ref{fig:multiple_story_lines}). The user starts out with an initial prompt to generate a number of potential story lines and decides which are worth continuing. The selected text outputs can then serve as the context for the next text prompt. Given the uncertainty of LLM this can be good way to spark the users imagination and explore different directions before deciding on a final story line. Another way multiple prompts can be organised can be seen in Fig. \ref{fig:prompt_view}, where prompts are structured in a separate view similar to the computational notebooks, e.g jupyter notebook.

\subsection{Supporting Users in Applying Prompts}
\begin{figure*}[t]
    \centering
    \includegraphics[width=0.5\textwidth]{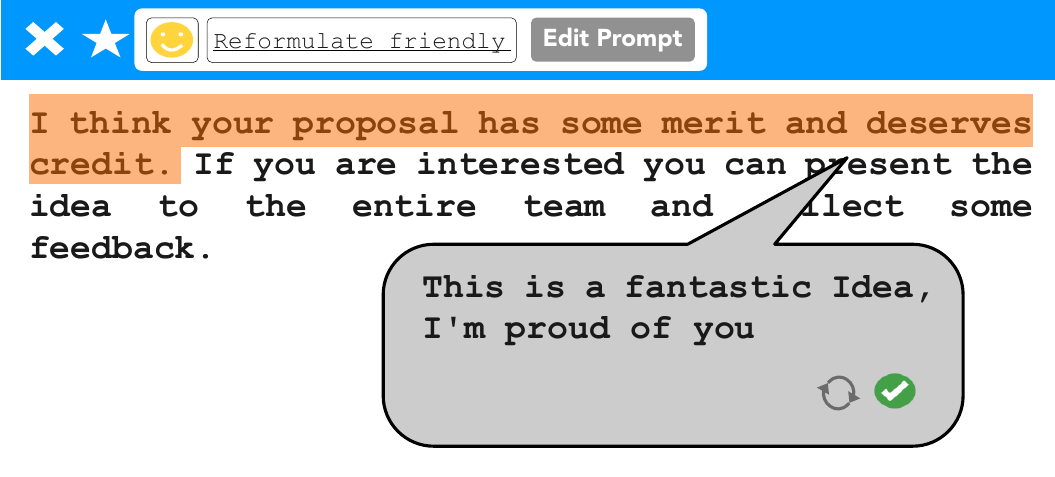}
    \caption{This GUI example supports users in applying prompts by enabling them to write and save prompts as tools in a toolbar. Users can then (re-)apply the text transformations defined by these prompts to text selections, such as the ``reformulate friendly'' tool here.}
    \label{fig:embedded_text_prompt}
\end{figure*}

Next to creating text prompts it is also important to think about how text prompts are embedded in a (known and existing) user interface and the interactions that apply or execute these prompts. %
In Fig. \ref{fig:embedded_text_prompt} we present a potential interaction flow where users assign a symbol to a reusable text prompt in a toolbar. Once such a defined symbol is selected the user can mark text to apply the underlying prompt, which brings up a dialog showing the generated output. %
More generally, stored prompts in the GUI could also be textual or, if possible, represented by a suitable symbol as illustrated in our example here.

\begin{figure*}[t]
    \centering
    \includegraphics[width=\textwidth]{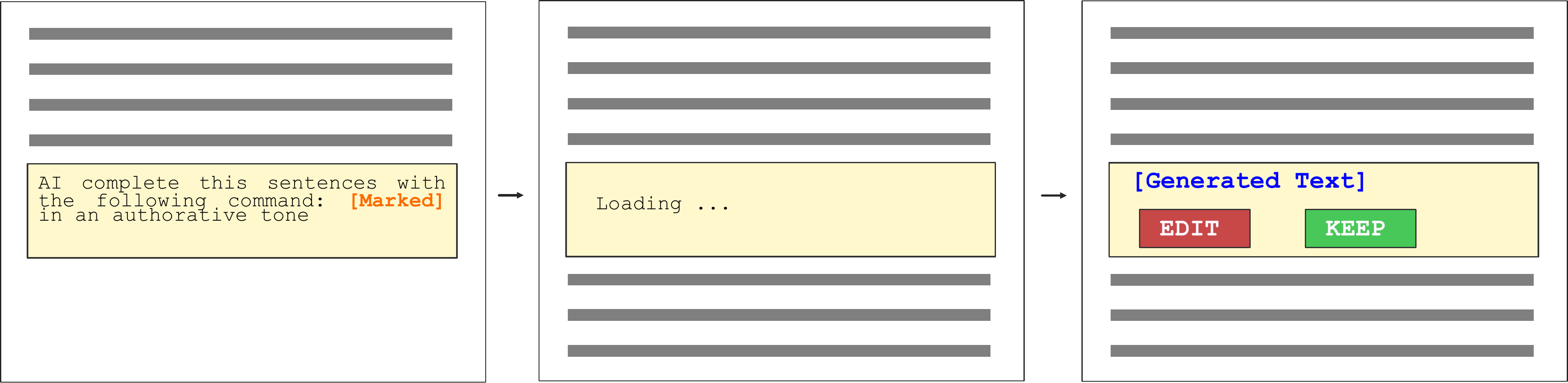}
    \caption{This example GUI supports asynchronous human-AI collaboration and in doing so also addresses potential delays of computationally costly prompts: The user can insert prompts into a text document anywhere, which trigger asynchronous calls to the LLM system, for example, to extend a part of the text. In the meantime, the user can continue working on another part of the text.}
    \label{fig:async_prompt}
\end{figure*}

As discussed, computational delay can be one of the challenges of designing effective interactions with text prompts. Thinking about how such delays can be incorporated in the design may help a user in applying prompts more creatively. For creative writing, one such opportunity could aim to maintain the writer's flow. For example, a user might have a short-term writer's block in which case it could be helpful to continue writing down another thought first, while letting the LLM finish the current paragraph (Fig. \ref{fig:async_prompt}). Once the results are ready the user will be notified and can return to edit this text passage.

\subsection{Representing Prompts in User Interfaces}
\begin{figure*}[t]
    \centering
    \includegraphics[width=0.5\textwidth]{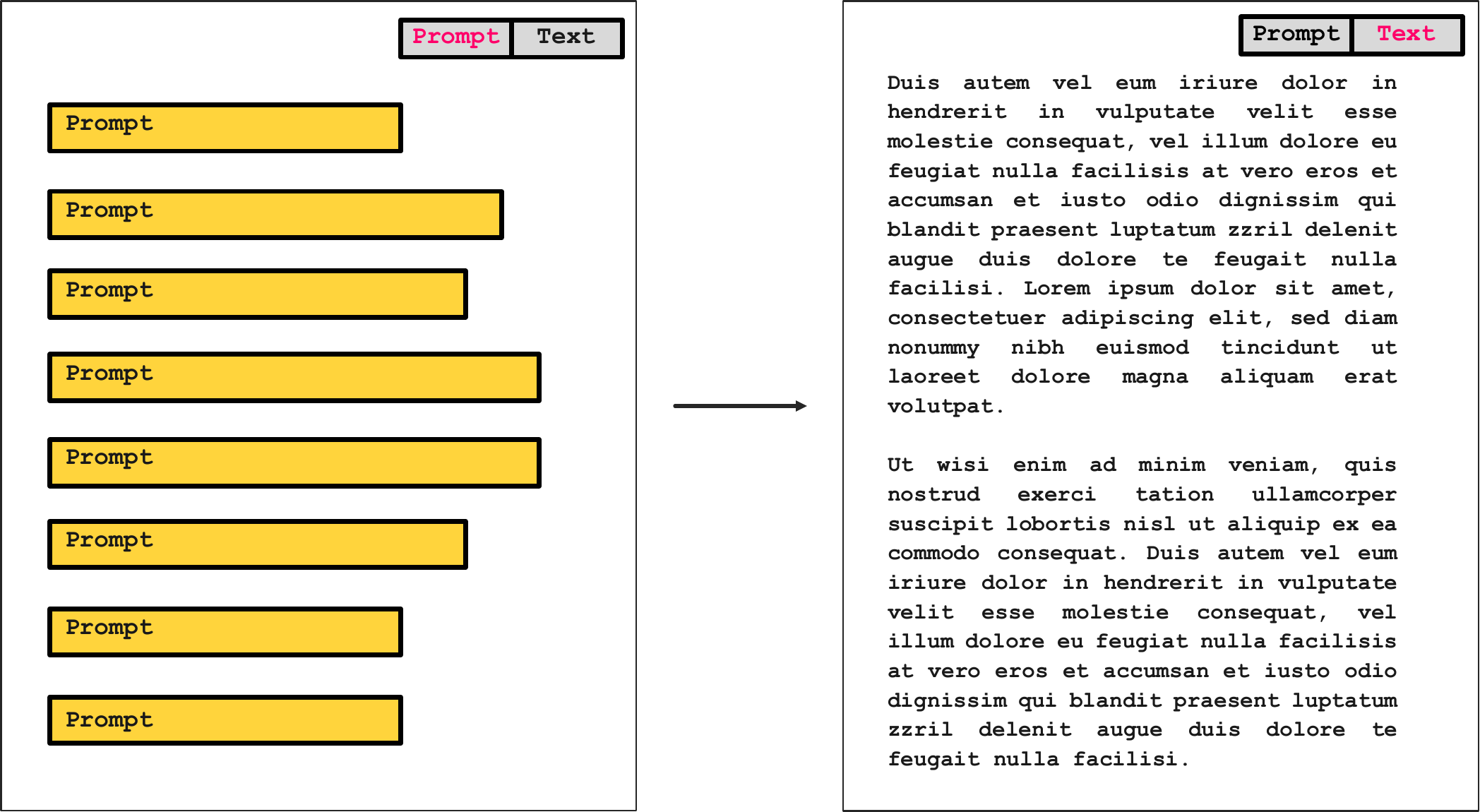}
    \caption{This GUI example treats the presentation of prompts similar to markdown, LaTeX or code editors: The GUI is split between prompt view and text view, allowing users to switch between editorial and evaluative tasks.}
    \label{fig:prompt_view}
\end{figure*}

Prompts also have to be represented in the user interface. One example is shown in Fig. \ref{fig:embedded_text_prompt} where a prompt is visualised as a symbol (smiley) and annotated to a text paragraph (highlighting, popup). Alternatively, prompts can also be organised along side a text. A simple split between a ``prompt'' and a ``text'' view allows users to switch between editorial and evaluative tasks (Fig. \ref{fig:prompt_view}). In this scenario, the prompt interface is similar to a hypertext or annotation interface, allowing users to define the structure and the creative annotations of the text. The resulting text view can serve as interface for light edits, but mainly to evaluate the generated output.

\section{Conclusion}

In this paper, we have argued to investigate prompting as a novel approach for human-AI interaction in creative tasks, motivated by promising opportunities around end-user programming of creative tools, creative expressiveness, and inspiration and feedback. %
Reflecting on the broader workshop theme, prompting  could democratise creative activities and domains, for example, by enabling users with no specific training to create personalised tools and more creatively work with text. Furthermore, users might be able to learn through the verbally expressive interactions and effects enabled by prompts and thereby improve their own creative skills. Prompting also extends creative possibilities -- not only how, but also what we can create -- by empowering individuals to flexibly and declaratively tap into the models' generative capabilities. In order to realise this potential we need adequate user interfaces for prompting. Future work should therefore explore design directions at the intersection of HCI and AI, such as the ones provided as starting points here.

\begin{acks}
This project is funded by the Bavarian State Ministry of Science and the Arts and coordinated by the Bavarian Research Institute for Digital Transformation (bidt).
\end{acks}

\bibliographystyle{ACM-Reference-Format}
\bibliography{bibliography}

\end{document}